\documentstyle[psfig,12pt,twoside,fleqn,espcrc1]{article}

\title{QCD perturbation theory at finite temperature/density and its application}

\author{M.H. Thoma\address{Institut f\"ur Theoretische Physik, Universit\"at Giessen,\\ 
        35392 Giessen, Germany}%
        \thanks{Heisenberg fellow}}
       
\begin{document}
\maketitle

\begin{abstract}
In order to describe properties of an equilibrated quark-gluon plasma, QCD at finite
temperature (and density) has to be considered. Besides lattice calculations, which can be
applied only to static quantities at zero density, perturbative QCD has been used. 
At finite temperature, however, serious problems such as infrared divergent and
gauge dependent results have been encountered. These difficulties can be (partially)
avoided if one starts from effective Green functions constructed by resumming a certain
class of diagrams (hard thermal loops). Within the last few years this improved 
perturbation theory (Braaten-Pisarski method) turned out to be a powerful tool
for computing interesting quantities of the quark-gluon plasma phase.

In the present talk a basic introduction to the Braaten-Pisarski method is provided
and its applications and limits are reviewed. In particular, damping rates, the energy 
loss of energetic partons, thermalization times, the viscosity of the quark-gluon
plasma, and the production of photons and dileptons are discussed.
\end{abstract}

\section{FINITE TEMPERATURE FIELD THEORY}

In order to detect a quark-gluon plasma (QGP) possibly produced in ultrarelativistic
heavy ion collisions, we are dependent on a thorough understanding of the properties of
a QGP and a prediction of signatures for the QGP formation. For this purpose we are forced
to use QCD at finite temperature and chemical potential, since the QGP is a relativistic
quantum systems of partons at finite temperature and baryon density. There are basically
two approaches to this problem. 

The first one is lattice QCD \cite{Ukawa}, which is 
able to describe a strong interacting system at all temperatures from 
below to above the phase transition from hadronic matter to a QGP. However, so far 
lattice QCD cannot be 
used to compute dynamical quantities such as particle productions from the QGP.
Also systems at finite density or out of equilibrium cannot be treated in this way.

The second approach is perturbative QCD at finite temperature. It is based on the fact that 
due to asymptotic 
freedom the effective temperature dependent strong coupling constant $\alpha _s(T)=g^2/4\pi $
becomes small at high temperature. For example, at $T=250 $ MeV the coupling constant
is expected to assume values of the order $\alpha _s =0.3$ - 0.5. To which extent perturbation
theory works for realistic situations, which can be realized in heavy ion collisions,
will be discussed at the end of this talk. 

Applying perturbative QCD at high temperatures static as well as dynamical quantities of a 
QGP can be considered. Also it is straightforward to extend this method to
finite quark chemical potential. Furthermore the application of perturbative QCD
to non-equilibrium situations, expected to take place in the first stage of a
relativistic heavy ion collision, is possible and the topic of current
investigations.   

An example for an important quantity calculable in thermal perturbative QCD is
the gluon self energy. To lowest order it is given by the diagrams shown in Fig.1.
Here we are using the following notation: $P\equiv (p_0,{\bf p})$ and $p\equiv |{\bf p}|$.

\begin{figure}
\leftline{\psfig{figure=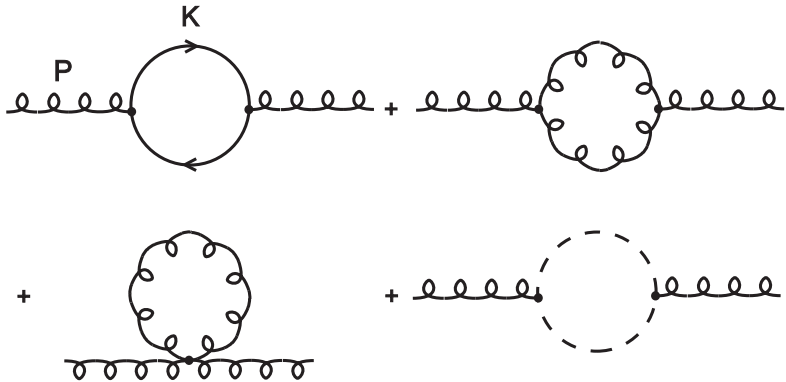,height=5cm}}
\vspace*{-0.5cm}
\caption{HTL gluon self energy containing quark, gluon, and ghost loops.}
\vspace*{-0.5cm}
\end{figure}

At finite temperature these diagrams can be evaluated by using either the imaginary
time (Matsubara) or the real time formalism \cite{Kapusta,LeBellac}. These
formalisms amount to a modification of the usual zero temperature Feynman rules 
in a way to include the thermal distribution functions of the partons. 
At finite temperature the diagrams of Fig.1, for example, can be evaluated only 
numerically \cite{Peshier}.
However, in the high temperature approximation a closed expression for the one-loop
gluon self energy can be found \cite{Klimov,Weldon}. The high temperature limit
can be shown to be equivalent to the so-called hard thermal loop (HTL) approximation,
introduced by Braaten and Pisarski \cite{Braaten1}. Here the diagrams of Fig.1
are computed by assuming the momenta of the internal lines to be much larger than
the ones of the external particles. In this way the following result has been obtained:
\begin{eqnarray}
\Pi_L(p_0,p) & = & -3m_g^2\> \left (1-\frac{p_0}{2p}\, 
\ln \frac{p_0+p}{p_0-p}\right ),\nonumber \\
\Pi_T(p_0,p) & = & \frac{3}{2} m_g^2\> \frac{p_0^2}{p^2}\> \left [1-\left 
(1-\frac{p^2}{p_0^2}\right )\, \frac{p_0}{2p}\, \ln \frac{p_0+p}{p_0-p}\right ],
\label{e1}
\end{eqnarray}
where $\Pi _L\equiv \Pi_{00}$ and $\Pi _T\equiv \Pi _{ij}\, (\delta _{ij}-p_ip_j/p^2)/2$
are the longitudinal and transverse components of the gluon self energy and
$m_g^2=g^2T^2(1+N_f/6)/3$ denotes the effective gluon mass, caused by the interaction with 
the partons of the plasma. ($N_f$ is the number of thermalized quark flavors in the QGP.)

It is interesting to note that the expressions given above for the gluon self energy are 
gauge independent in contrast to the complete one-loop expressions. 

An effective gluon propagator can be constructed by resumming 
the HTL gluon self energy within the Dyson-Schwinger equation,
%
leading to 
\begin{eqnarray}
D_L^*(P) & = & \frac{1}{p^2-\Pi _L(p_0,p)},\nonumber \\
D_T^*(P) & = & \frac{1}{P^2-\Pi _T(p_0,p)}
\label{e3}
\end{eqnarray}
for the longitudinal and transverse part of the resummed gluon propagator in Coulomb gauge.
This propagator describes the propagation of collective plasma modes.
The dispersion relation of these plasma modes given by the poles of the 
propagator is shown in Fig.2. There are two branches, one for the longitudinal
plasma mode, also called plasmon, and one for the transverse modes. Both
start at the same energy $m_g$ (plasma frequency) for zero momentum and approach the
dispersion relation of a bare gluon at large momenta. 

\begin{figure}
\leftline{\psfig{figure=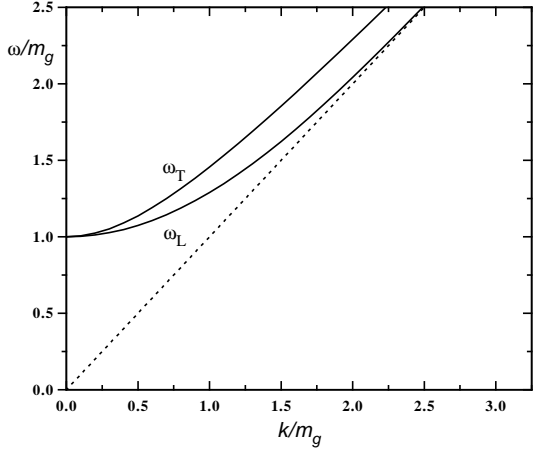,height=6cm}}
\vspace*{-0.5cm}
\caption{Gluon dispersion relations in a QGP.}
\vspace*{-0.5cm}
\end{figure}

The HTL gluon self energy exhibits an imaginary part for $p_0^2<p^2$ leading to Landau 
damping below the light cone. The HTL plasma modes, however, are not damped as they
are situated entirely above the light cone. After all the Landau damping has some interesting
consequences as we will discuss below. 

In the static limit, $p_0=0$, the longitudinal HTL gluon self energy reduces to $m_D^2=
3m_g^2$. As a consequence the resummed longitudinal propagator shows Debye screening due to the
presence of color charges in the plasma. On the other hand, there is no static
magnetic screening in the HTL approximation since $\Pi _T(p_0=0)=0$.

\section{HTL RESUMMATION TECHNIQUE}
 
Restricting to bare propagators perturbative QCD can lead to infrared divergent and
gauge dependent results for physical quantities. A famous example is the damping 
rate $\gamma $ of a gluon plasma mode at rest, 
%
%
which follows from the imaginary part of the gluon self energy. Using only bare propagators
and vertices
$\gamma = a\> g^2T/(8\pi )$
has been found with $a=1$ in Coulomb or temporal gauge \cite{Kajantie} and $a=-5$
in Feynman gauge \cite{Lopez}. (A negative damping rate corresponds to an unstable QGP.)

The reason for this unphysical behavior can be traced back to the fact that perturbative
QCD at finite temperature is incomplete, i.e., higher order diagrams (multi-loop diagrams)
can contribute to lower order in the coupling constant \cite{Pisarski2}. In order to
arrive at a consistent expansion at finite temperature, these diagrams, which are
exactly the HTL diagrams, have to be resummed into effective propagators and vertices.
Besides the effective gluon propagator in (\ref{e3}) there is a resummed quark 
propagator containing the quark self in the HTL approximation. Furthermore,
HTL vertices show up in gauge theories as they are related to the HTL self energies 
by Ward identities.   


Now, assuming that the weak coupling limit, $\alpha _s\ll 1$, holds, we see e.g from
(\ref{e1}) and (\ref{e3}) that bare propagators are sufficient as long as the momentum is hard,
$p^2\sim T^2$. However, for soft momenta, $p^2\sim \alpha _sT^2$, a resummed 
propagator has to be used. Effective vertices are only necessary if all external legs 
of the vertex are soft. This is the basic idea of the HTL resummation technique
developed by Braaten and Pisarski \cite{Braaten1}. In this way consistent results
are obtained, i.e., gauge independent results which are complete to leading order 
in the coupling constant. At the same time the infrared behavior of these quantities is
improved due to Debye screening in the resummed gluon propagator.  

In the case of the plasmon damping rate at zero momentum the Braaten-Pisarski method 
%
leads to the gauge independent result $\gamma =6.636\> g^2T/(8\pi )$ \cite{Braaten2}.

The HTL resummation technique can easily be extended to finite baryon density. 
For example in the case of the gluon propagator one simply has 
to replace the effective gluon mass in (\ref{e1}) by \cite{Vija}
\begin{equation}
m_g^2=\frac{g^2T^2}{3}\> \left [1+\frac{1}{6}
\left (N_f+\frac{3}{\pi ^2}\sum_f \frac{\mu _f^2}{T ^2}\right )\right ],
\label{e8}
\end{equation}
where $\mu _f$ is the chemical potential of the quarks with flavor $f$.

Finally, the Braaten-Pisarski method can also be modified to non-equilibrium situations.
Basically, one has to replace the thermal distribution functions by non-equilibrium
ones (Wigner functions), although some subtleties such as pinch singularities have
to be considered \cite{Baier1,Carrington}.

\section{APPLICATIONS}

In this section I will discuss the applications of the HTL resummation technique to
possible signatures and other interesting observables of a QGP \cite{Thoma1}.

\subsection{Damping rates of energetic partons}

The damping rates of quarks and gluons with energies $E\gg T$ in a QGP are one of the most 
discussed applications of the Braaten-Pisarski method (for references see \cite{Thoma1}).

Within naive perturbation theory, i.e. using only bare Green functions, the damping rate
of a hard quark follows from the left diagram of Fig.3, which corresponds to scattering 
of the 
energetic quark off the thermal partons via the exchange of a gluon. The matrix 
element of the scattering diagram is related to the imaginary part of a two-loop 
quark self energy
as can be seen by cutting through the self energy. From these diagrams
we expect that the damping rate is of order $g^4$. Owing to the exchange of a bare
gluon the damping rate in naive perturbation theory turns out to be quadratically
infrared divergent. 

\begin{figure}
\leftline{\psfig{figure=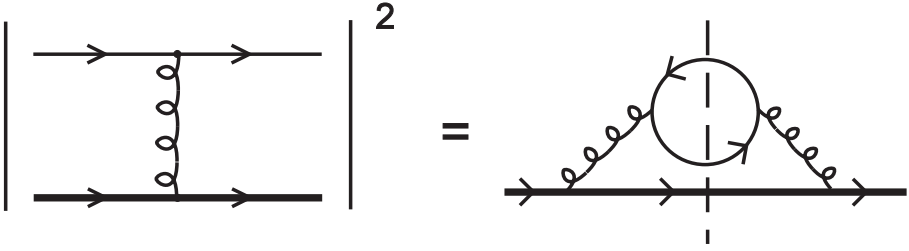,width=12cm}}
\vspace*{-0.5cm}
\caption{Diagrams for the damping rate within naive perturbation theory.}
\vspace*{-0.5cm}
\end{figure}

Using the resummed perturbation theory, the damping to lowest order is derived
from the imaginary part of the quark self energy shown in Fig.4 containing 
a resummed gluon propagator. The imaginary part of this diagram is due to 
the imaginary part of the HTL gluon self energy of the resummed propagator.  
Therefore the damping mechanism is caused by the exchange of a virtual plasma mode, 
which is damped below the light cone (virtual Landau damping). 

\begin{figure}[h]
\leftline{\psfig{figure=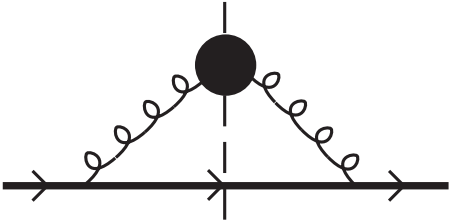,height=3cm}}
\vspace*{-0.5cm}
\caption{Diagram for the damping rate within resummed perturbation theory.}
\vspace*{-0.5cm}
\end{figure}


Computing the damping rate from Fig.4 leads to a logarithmic infrared singularity,
which comes from the exchange of the transverse plasma mode containing no static
magnetic screening. Hence, in the case of the damping rate of moving partons
the HTL resummation method is not sufficient to remove all infrared singularities.
Assuming an infrared cutoff of the order $g^2 T$, which could be provided either by
a non-perturbative magnetic mass or by the finite width of the hard quark in the QGP,
the following result has been obtained \cite{Thoma1}:
\begin{equation}
\gamma _q =\frac{g^2T}{3\pi }\> \ln \frac{1}{g}.
\label{e9}
\end{equation}
Note that the leading order contribution to the damping rate is of order $g^2$ in
contrast to naive expectation. This anomalous large damping \cite{Lebedev}
is caused by the use of the resummed gluon propagator, which contains the
HTL gluon self energy of order $g^2$ in the denominator.

A possible physical implication of this damping rate is the color conductivity, which
determines the relaxation of the QGP in the case of a small deviation from an equilibrium
color distribution. It   
is given by  $\sigma _c=4m_g^2/\gamma _g$ \cite{Selikhov,Heiselberg}, where $\gamma _g$ 
is now the damping rate of a thermal gluon, which follows from (\ref{e9})
by multiplying by a color factor 9/4. 

\subsection{Energy loss of energetic partons}

The energy loss of an energetic parton in the QGP is related to jet quenching in 
ultrarelativistic heavy ion collisions at RHIC and LHC, where jets are expected
to arise from high energy partons coming from the primary hard parton collisions.
Partons with a large transverse momentum have to propagate through the fireball  
thereby losing their energy. The amount of jet quenching, produced in this way, may
depend on the phase of the fireball and might be used as a signature for the QGP
formation therefore \cite{Gyulassy1}. Another application of the energy loss of heavy 
quarks is the suppression of dileptons with large invariant masses from the decay of
charm mesons \cite{Lin,Mustafa}.

To lowest order there are two contributions to the energy loss of a parton in the QGP,
one coming from elastic scattering (collisional energy loss) and the other one
from gluon bremsstrahlung (radiative energy loss). The collisional energy loss
can be derived by dividing the energy transfer per collision $\Delta E$ by the mean free 
path of
the parton. The mean free path due to elastic scattering is proportional to the inverse
of the damping rate times the velocity $v$ of the parton. Therefore the collisional energy 
loss of a quark, for example, can be calculated from \cite{Braaten3}
\begin{equation}
\frac{dE}{dx}=\frac{2}{v}\> \int d\gamma _q\> \Delta E.
\label{e10}
\end{equation}
%
Owing to the 
additional factor $\Delta E$ under the integral defining the damping rate
the collisional energy loss is only logarithmically 
infrared divergent
within naive perturbation theory and finite using the HTL resummation technique.
Using a resummed gluon propagator (Fig.4) for soft momenta of the exchanged gluon and
a bare gluon propagator for hard momenta (Fig.3), we end up with an extension of the Bethe-Bloch 
formula to the case of a QGP \cite{Braaten4,Thoma3}:
\begin{equation}
\frac {dE}{dx} = \frac{16\pi }{9} \, \alpha _s^2\, T^2\> \ln \frac{9E}{16\pi 
\alpha _sT}. 
\label{e11}
\end{equation}
Note that the result is of order $g^4$ in contrast to the damping rate, which is caused
by the reduction of the infrared divergence by the energy transfer factor $\Delta E$.
Inserting typical values for $\alpha _s= 0.3$ - 0.5, $T= 200$ - 300 MeV and $E= 10$ - 50 GeV,
we find an energy loss of the order $dE/dx\simeq 0.5$ - 1.5 GeV/fm. In the case of 
an energetic gluon we have to multiply this result by 9/4 again. In Fig.5 
the collisional
energy loss of charm and bottom quarks is shown as a function of the quark momentum 
for various values of the quark chemical potential \cite{Vija}.

\begin{figure}
\leftline{\psfig{figure=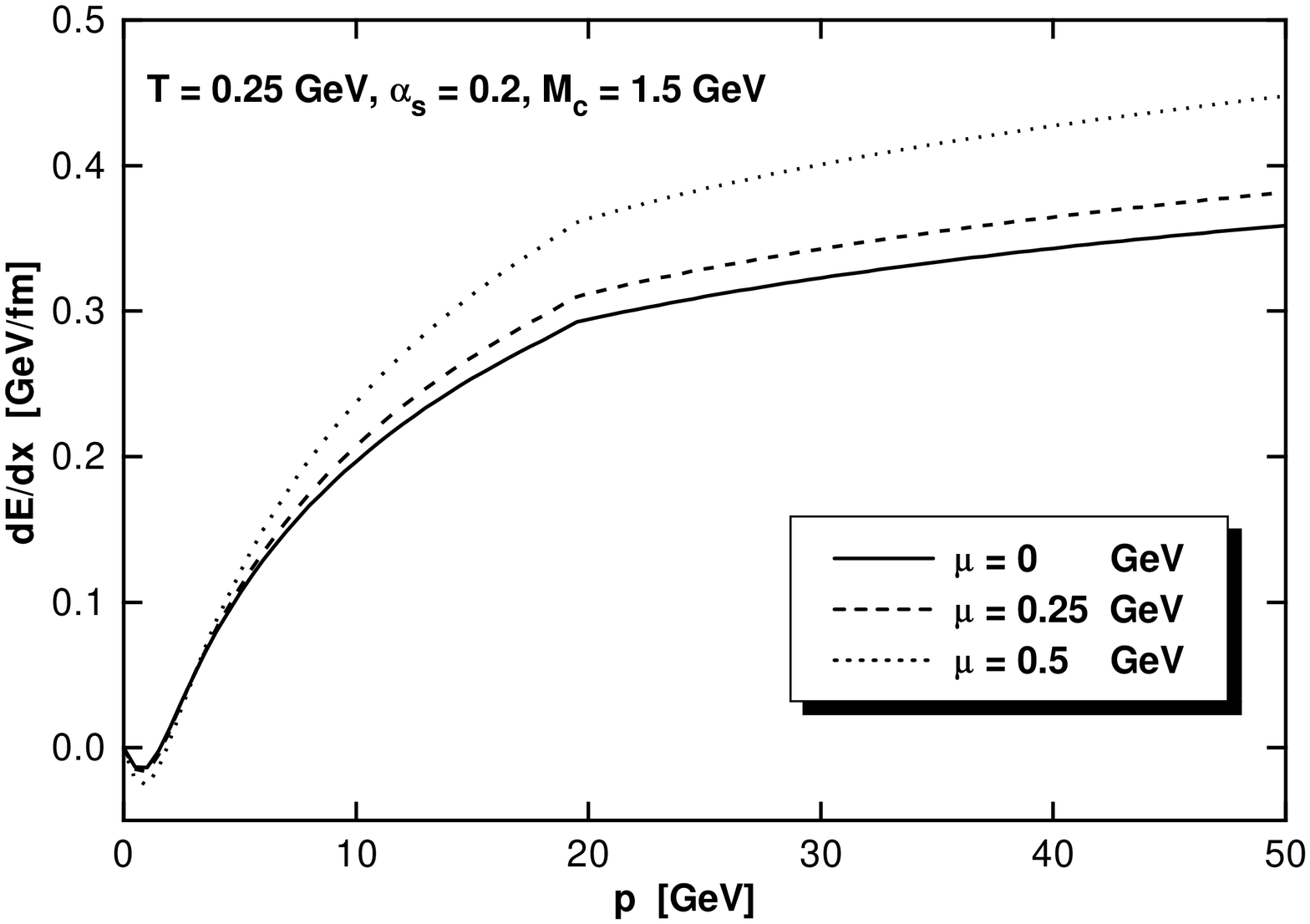,height=5cm}\hfill {\psfig{figure=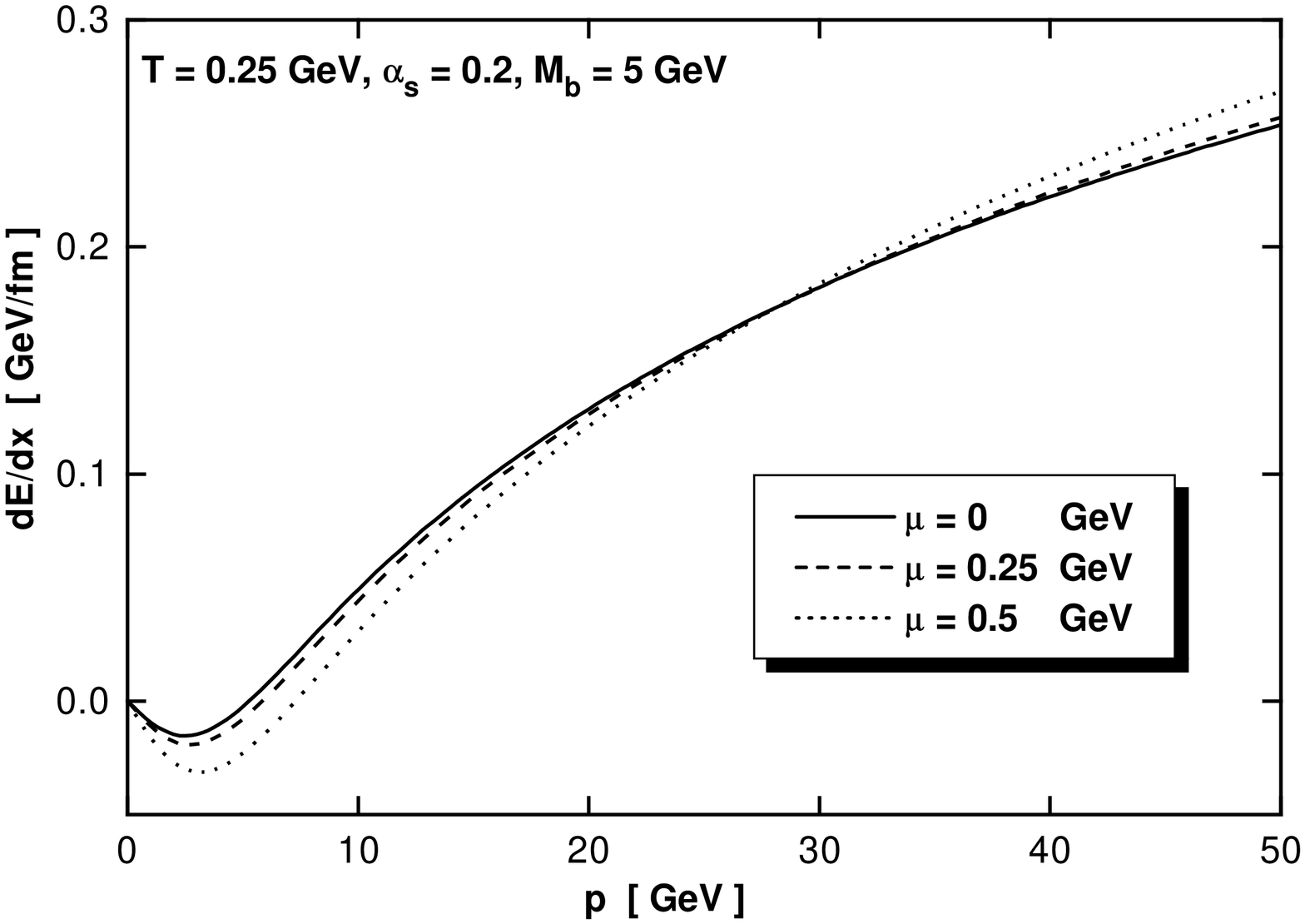,height=5cm}}}
\vspace*{-0.5cm}
\caption{Collisional energy loss of charm and bottom quarks.}
\vspace*{-0.5cm}
\end{figure}


The calculation of the radiative energy loss due to gluon bremsstrahlung is much more
involved.
For example, the suppression of the gluon emission due to multiple scattering 
(Landau-Pomeranchuk effect) has to be taken into account. Estimates of the radiative
energy loss indicate that it dominates over the collisional one typically by a factor of
five \cite{Gyulassy2,Baier2}. 


\subsection{Thermalization times and viscosity}

In order to estimate thermalization times we have to consider the momentum relaxation
in the case of a small deviation from the equilibrium momentum distribution. The
momentum relaxation or transport rate is given by
\begin{equation}
\gamma _{trans}=\int d\gamma \> \frac{\sin ^2\theta }{2},
\label{e12}
\end{equation}
where the transport factor $(\sin ^2\theta)/2$ takes into account that collinear
scattering angles $\theta $ do not contribute to the momentum relaxation. 

As in the case of the collisional energy loss the transport factor reduces the 
degree of the infrared singularity compared to the damping rate, yielding a finite 
result within the resummed perturbation theory. Calculating the transport rate analogously 
to the collisional energy loss, the inverse thermalization times, $\tau ^{-1}\equiv 2\, 
\gamma _{trans}$, for quarks and gluons read \cite{Thoma4}
\begin{eqnarray}
\tau _q^{-1} & = & 2.5\> \alpha _s^2T\> \ln \frac{0.21}{\alpha _s}, \nonumber \\
\tau _g^{-1} & = & 6.6\> \alpha _s^2T\> \ln \frac{0.19}{\alpha _s}.
\label{e13}
\end{eqnarray}
 
This result suggest that gluons thermalize about three times faster than quarks.
Hence one might expect that first 
the gluon component of the primordial parton gas thermalizes and the thermalization
of the quark component sets in later \cite{Shuryak}. 

However, the results (\ref{e13}) become unphysical, i.e. negative,  for realistic values of 
the coupling constant, $\alpha _s>0.2$, indicating the break down of perturbation theory. 
This behavior can be traced back to the use of thermal energies $E\simeq 3T$ for the
partons in (\ref{e13}), whereas for high energies $E\gg T$, as in the case of the
energy loss (\ref{e11}), reasonable results are found.
 
The shear viscosity coefficient of the QGP is closely related to the transport rate.
Using the relaxation time approximation it is given by $\eta =(1/15)\, \epsilon\, 
\gamma _{trans}^{-1}$ \cite{Daniel}, where $\epsilon $ is the energy density of the plasma.
The viscosity of the QGP is the sum of the viscosities of the gluon and the quark component   
\cite{Thoma4}
\begin{equation}
\eta=\frac{T^3}{\alpha _s^2}\> \left [\frac{0.11}{\ln (0.19/\alpha _s)}+
\frac{0.37}{\ln (0.21/\alpha _s)}\right ].
\label{e14}
\end{equation}

In Fig.6 the viscosity coefficient of the QGP divided by $T^3$ is shown as a function
of the coupling constant. The dashed curve corresponds to an estimate by Baym et al.
\cite{Baym}, which is not based on the HTL resummation method. The dotted line indicates 
the upper limit for the validity of the Navier-Stokes equations \cite{Daniel}.
This suggests that dissipation effects are important in the QGP and should not be neglected
in hydrodynamical calculations.

\begin{figure}[h]
\leftline{\psfig{figure=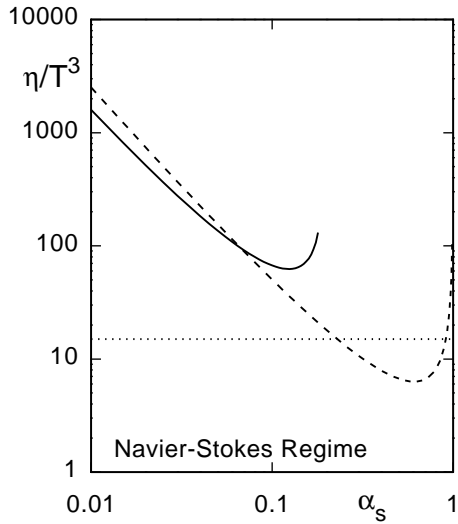,height=7cm}}
\vspace*{-0.5cm}
\caption{Viscosity coefficient of the QGP.}
\vspace*{-0.5cm}
\end{figure}

\subsection{Photon production}

The thermal emission of photons and dileptons from the QGP is one of the proposed 
signatures for
the deconfinement transition \cite{Ruuskanen}. For this purpose one has to know
the photon production rate from the QGP as well as from a hadron gas. 

Let us first consider hard photons with an energy $E\gg T$. To lowest order 
these photons are produced by the diagrams of Fig.7, i.e. Compton scattering and
quark annihilation. 

\begin{figure}[h]
\leftline{\psfig{figure=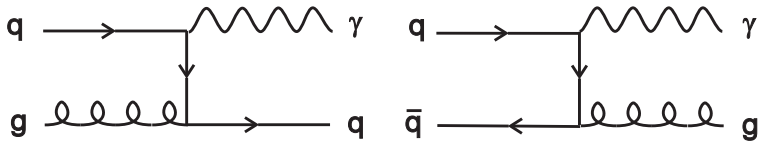,width=10cm}}
\vspace*{-0.5cm}
\caption{Photon production diagrams in a QGP.}
\vspace*{-0.5cm}
\end{figure}

Assuming the quarks to be massless, the diagrams of Fig.7 lead to a logarithmically
divergent production rate. Using the resummed quark propagator   
for soft momenta, on the
other hand, a finite result is obtained. (Taking into account the bare quark masses leads
also to a finite result, which is, however, much too large because the effective
quark masses of the order $gT$ provide a much larger infrared cutoff.) 

The photon production rate from a QGP has been derived using a resummed propagator for soft 
and a bare one for hard momenta of the exchanged quark \cite{Kapusta2,Baier3}. In the case 
of a finite quark chemical $\mu $ potential the following result has been found \cite{Traxler}:
\begin{equation}
E\frac{dR}{d^3p}=\frac{5\alpha \alpha _s}{18\pi^2}\> e^{-E/T}\> \left (T^2+
\frac{\mu ^2}{\pi ^2}\right )\> \ln \frac{0.13E}{\alpha _sT},
\label{e15}
\end{equation}
where $\alpha $ is the QED fine structure constant. 


This rate is of the same order \cite{Kapusta2} or even two to three times smaller 
\cite{Shuryak2,Haglin} than the estimate of the photon production rate from a 
hadronic gas at the same 
temperature. In order to calculate the photon spectrum the space-time evolution of the 
fireball has to be considered. Using (\ref{e14}) in a hydrodynamical model containing 
a QGP, a mixed, and a hadronic phase a photon spectrum for SPS ({\it S+Au}) is found \cite{Arbex}
which agrees with the experimental upper limit for the direct photon production
\cite{WA80}.

The production rate of soft photons ($E\ll T$) can be calculated from the imaginary part of the
one-loop photon self energy containing effective propagators and vertices.
Owing to the singularity of the resummed Green functions near the light cone, e.g.
for $p_0 \rightarrow p$ in (\ref{e1}), the soft production rate of on-shell photons
exhibits a collinear singularity \cite{Baier4}.

\subsection{Dilepton production}

As a last example we consider the production rate of dileptons from the QGP. First we focus 
on hard dileptons, where the energy $E$ of the virtual photon is much larger than the
temperature. The lowest order contribution of order  $\alpha ^2$ comes from the Born term, 
i.e. the annihilation of a quark-antiquark pair into a virtual photon.  

\begin{figure}[h]
\leftline{\psfig{figure=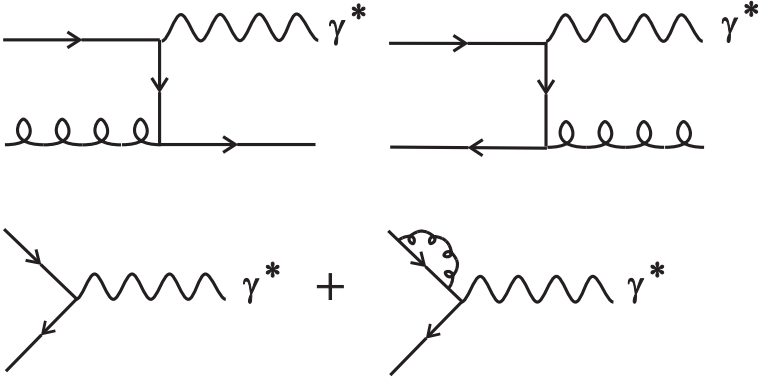,height=5cm}}
\vspace*{-0.5cm}
\caption{$\alpha _s$-corrections of the hard dilepton production in the QGP.
The sum of the diagrams in the lower line denote the interference term of these graphs.}
\vspace*{-0.5cm}
\end{figure}

For invariant masses $M^2\equiv E^2-p^2\rightarrow 0$ the Born contribution 
vanishes. Therefore one has to consider the $\alpha _s$-corrections to the dilepton
rate for $M{\buildrel <\over \sim}T$.  This contribution follows from the diagrams of Fig.8, 
where besides the graphs leading to the real photon production one has to take into account 
the interference term of the Born diagram with an annihilation diagram containing a quark 
self energy insertion. Adding up these contributions the dilepton production rate turns
out to be infrared finite even for massless quarks in contrast to the photon rate due to   
a cancellation of infrared divergences of the interference term and the other diagrams of 
Fig.8 \cite{Altherr}. However, although the result is finite, one has to use a resummed
quark propagator analogously to the real photon case in order to obtain a consistent result.
Proceeding similarly to the photon production rate, the following result has been found 
\cite{Thoma5}
\begin{equation}
\frac{dR}{d^4xd^4p}=\frac{10}{27\pi ^3}\> \alpha ^2\alpha _s\> \frac{T^2}{M^2}\> 
e^{-E/T}\> \left (\ln \frac{T(m_q+k_{min})}{m_q^2}+C\right ),
\label{e16}
\end{equation}
where $m_q^2=g^2T^2/6$ is the effective quark mass, $k_{min}\simeq |Em_q^2/M^2-M^2/(4E)|$, and
$C\simeq -0.6$. This result is shown in Fig.9 as a function of the photon energy $E$ 
and the invariant mass $M$ for 
$\alpha _s=0.3$, $T=300$ MeV. For comparison the result found by
Altherr and Ruuskanen \cite{Altherr} calculated with bare propagators and the Born 
contribution are shown. For smaller invariant masses the $\alpha _s$-corrections dominate over 
the Born term, which is independent of $M$. For example, for $M=100$ MeV and $E=3$ GeV the 
$\alpha _s$-correction exceeds the Born contribution by a factor of about 20.  

\begin{figure}
\leftline{\psfig{figure=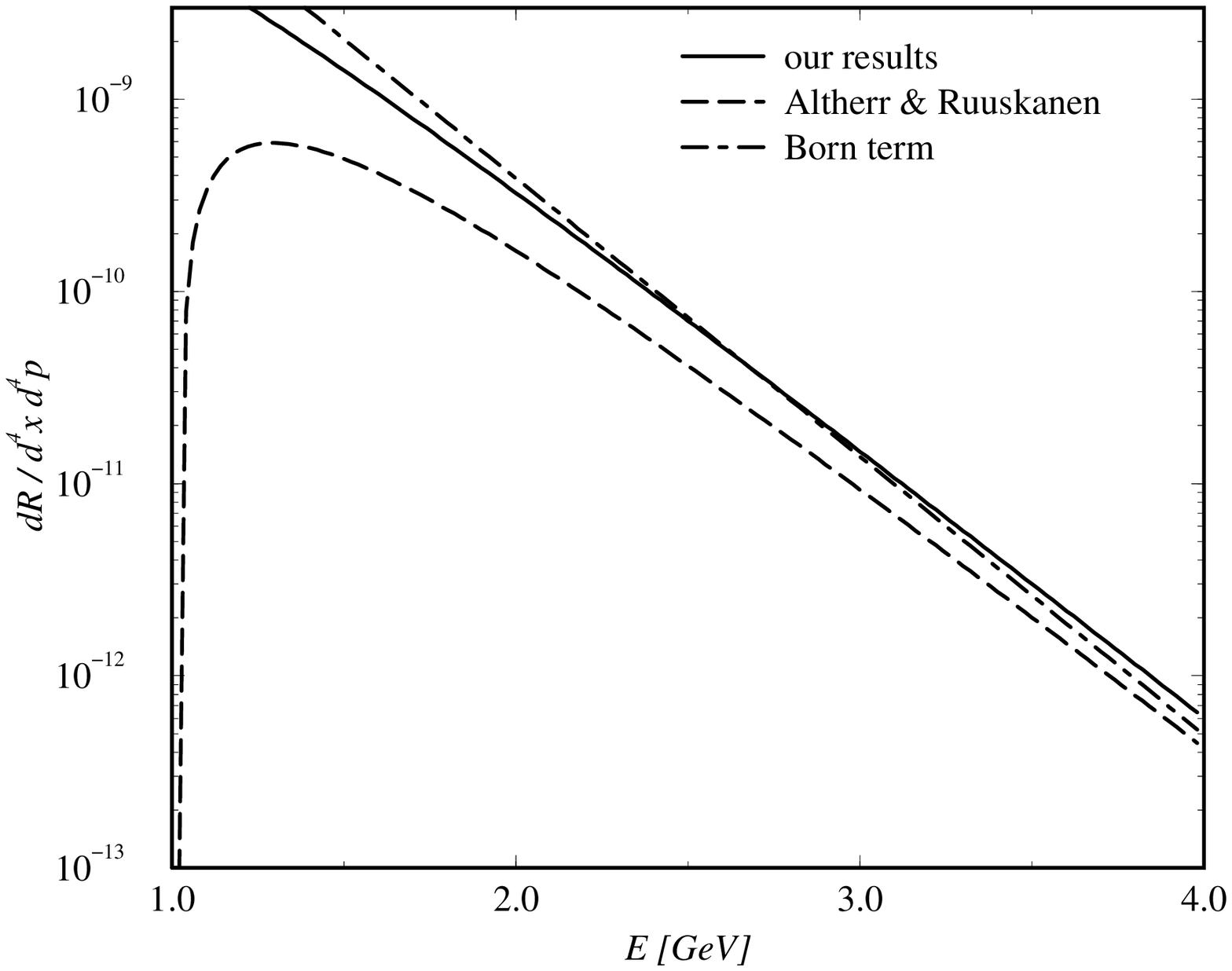,height=7cm}\hfill {\psfig{figure=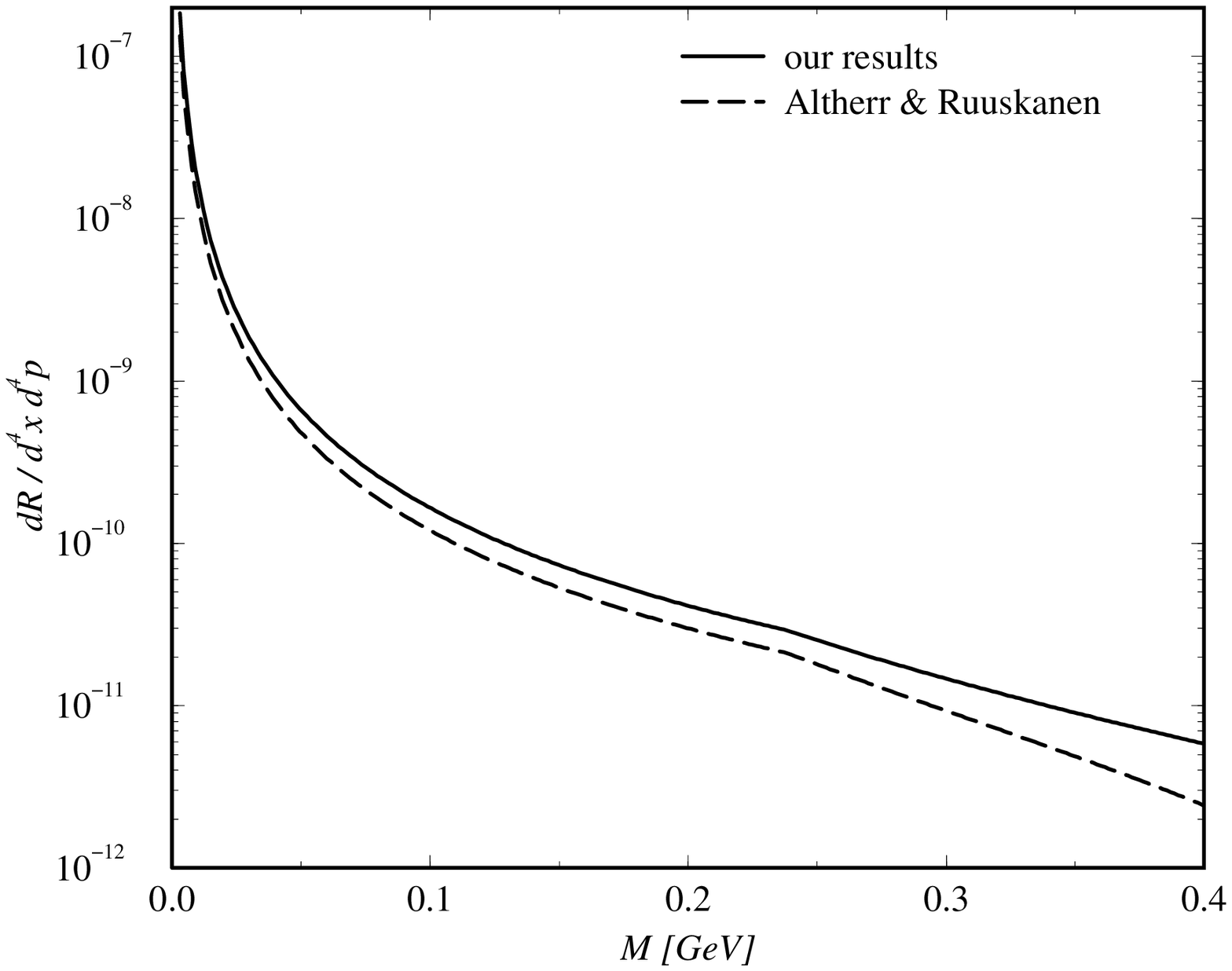,height=7cm}}}
\vspace*{-1cm}
\caption{Hard dilepton production rate from the QGP ($\alpha _s=0.3$, 
$T=300$ MeV) as a function of $E$ for $M=300$ MeV (left figure) and as a function of $M$ for 
$E=3$ GeV (right figure).}
\vspace*{-0.5cm}
\end{figure}

The production rate of soft dileptons can de derived from the same photon self energy as
the soft photon production \cite{Braaten5,Wong}. However, due to the off-shellness of 
the photons there is no collinear infrared singularity in this case. The soft photon 
production shows an interesting structure as a function of the photon energy caused by the
dispersion relation of the soft quark modes in the QGP \cite{Braaten5}.

\section{CONCLUSIONS}

Using perturbative QCD at finite temperature static as well as dynamical properties
of a QGP can be calculated.
This method can be extended in a straightforward manner
to finite baryon density, i.e. finite quark chemical potential, and even to
non-equilibrium situations using the Keldysh formalism.  
Perturbative QCD is expected to work at high temperatures, where the effective
strong coupling constant becomes small due to asymptotic freedom. However,
using only bare propagators and vertices infrared divergent and gauge dependent results
are obtained even for physical quantities. Using, however, HTL resummed Green functions 
for soft momenta of the order $gT$ and smaller, consistent results can be derived, i.e.
gauge independent results complete to leading order in the coupling constant. In many
cases all infrared singularities are removed to leading order in the coupling constant
in this way. At the same time important medium effects of the QGP, such as effective 
parton masses, Debye screening, and Landau damping, are taken into account.

The HTL resummation technique has been applied to the following observables of the QGP:

\smallskip

\noindent
1. Damping rates of quarks and gluons in the QGP, which are related to color conductivity of 
the QGP.  

\noindent  
2. The energy loss of energetic partons in the QGP, which determines the amount of jet 
quenching at RHIC and LHC and the suppression of high invariant mass dileptons from
the decay of charm mesons.

\noindent
3. Thermalization times of quarks and gluons, which show that gluons equilibrate faster
than quarks.

\noindent
4. The viscosity of the QGP indicating the significance of dissipative effects in the QGP.

\noindent
5. Photon and dilepton production rates, which serve as a promising signature for the QGP
formation.

\smallskip

There are two open problems concerning the application of the HTL resummation technique. The 
first one is the weak coupling limit ($\alpha _s\ll 1$), which had been assumed in the
derivation of the HTL method. For realistic values of the coupling constant the HTL
calculations to lowest order fail if the energy of the quantity under consideration 
is of the order of the temperature or smaller. Examples are the Debye mass, which is about
a factor of three smaller than the one obtained non-perturbatively \cite{Karsch,Kajantie2},
and the thermalization times and the viscosity discussed here, which become negative for 
$\alpha _s>0.2$. For large energies ($E\gg T$), on the other hand,
the HTL resummation technique produces reasonable results, such as the collisional energy    
loss of energetic partons and the production rates of hard photons and dileptons.

The second problem concerns infrared singularities. Although the inclusion of Debye screening
using the HTL method reduces the infrared singularities to a large extent
compared to naive perturbation theory, two classes of singularities still survive. 
The first one is caused by the absence of static magnetic screening in the HTL resummed
gluon propagator. Either a magnetic screening mass or a finite width of the partons
in the QGP can cure this problems. Magnetic screening masses require the use of 
non-perturbative methods \cite{Linde}, while a finite width of the thermal partons can be
achieved by a Bloch-Nordsieck resummation \cite{Blaizot}. The second cause 
for possible infrared singularities is the divergence of the resummed Green functions
at the light cone. In the case of the soft photon production this leads to a collinear
infrared singularity. In order to remove this singularity an improved HTL resummation
scheme has been proposed \cite{Rebhan2}.

\end{document}